\documentclass[aps,prl,twocolumn,showpacs,superscriptaddress]{revtex4}

\usepackage{graphicx}
\usepackage{amsmath}
\usepackage{amsfonts}
\usepackage[english]{babel}
\usepackage{bm}    

\def\reff#1{(\ref{#1})}
\newcommand{\be}{\begin{equation}}
\newcommand{\ee}{\end{equation}}
\newcommand{\<}{\langle}
\renewcommand{\>}{\rangle}

\def\spose#1{\hbox to 0pt{#1\hss}}
\def\ltapprox{\mathrel{\spose{\lower 3pt\hbox{$\mathchar"218$}}
 \raise 2.0pt\hbox{$\mathchar"13C$}}}
\def\gtapprox{\mathrel{\spose{\lower 3pt\hbox{$\mathchar"218$}}
 \raise 2.0pt\hbox{$\mathchar"13E$}}}

\newcommand{\scrd}{{\cal D}}

\newcommand{\scrh}{{\cal H}}

\def\bsigma{{\boldsymbol{\sigma}}}
\def\vecbsigma{{\vec{\boldsymbol{\sigma}}}}
\def\bpi{{\boldsymbol{\pi}}}
\def\bGamma{{\boldsymbol{\Gamma}}}

\def\psibar{{\bar{\psi}}}

\def\R{{\mathbb R}}

\newenvironment{scarray}{
          \textfont0=\scriptfont0
          \scriptfont0=\scriptscriptfont0
          \textfont1=\scriptfont1
          \scriptfont1=\scriptscriptfont1
          \textfont2=\scriptfont2
          \scriptfont2=\scriptscriptfont2
          \textfont3=\scriptfont3
          \scriptfont3=\scriptscriptfont3
        
        \begin{array}{c}}{\end{array}}

\begin{document}

\title{Fermionic field theory for trees and forests}

\author{Sergio Caracciolo}
\affiliation{Universit\`a degli Studi di Milano -- Dipartimento di Fisica
      and INFN, via Celoria 16, I-20133 Milano, Italy}
\author{Jesper Lykke Jacobsen}
\affiliation{LPTMS, Universit\'e Paris-Sud, B\^atiment 100,
      F-91405 Orsay, France}
\author{Hubert Saleur}
\affiliation{Service de Physique Th\'eorique, CEA Saclay,
      F-91191 Gif-sur-Yvette, France}
\affiliation{Department of Physics and Astronomy,
   University of Southern California, Los Angeles, CA 90089, USA}
\author{Alan D. Sokal}
\affiliation{Department of Physics, New York University,
      4 Washington Place, New York, NY 10003, USA}
\author{Andrea Sportiello}
\affiliation{Universit\`a degli Studi di Milano -- Dipartimento di Fisica
      and INFN, via Celoria 16, I-20133 Milano, Italy}

\date{9 March 2004; revised 8 June 2004}

\begin{abstract}
We prove a generalization of Kirchhoff's matrix-tree theorem
in which a large class of combinatorial objects are represented
by non-Gaussian Grassmann integrals.
As a special case, we show that unrooted spanning forests,
which arise as a $q \to 0$ limit of the Potts model,
can be represented by a Grassmann theory involving a Gaussian term
and a particular bilocal four-fermion term.
We show that this latter model can be mapped,
to all orders in perturbation theory,
onto the $N$-vector model at $N=-1$ or, equivalently,
onto the $\sigma$-model taking values in the unit supersphere in $\R^{1|2}$.
It follows that, in two dimensions,
this fermionic model is perturbatively asymptotically free.
\end{abstract}

\pacs{05.50.+q, 02.10.Ox, 11.10.Hi, 11.10.Kk}

\keywords{Kirchhoff theorem, matrix-tree theorem, spanning forests,
Potts model, Grassmann algebra, $N$-vector model, $\sigma$-model,
supersymmetry, asymptotic freedom.}

\maketitle


Kirchhoff's matrix-tree theorem \cite{Kirchhoff_et_al}
and its generalizations \cite{Chaiken_et_al},
which express the generating polynomials of spanning trees
and rooted spanning forests in a graph
as determinants associated to the graph's Laplacian matrix,
play a central role in electrical circuit theory \cite{Balabanian_69}
and in certain exactly-soluble models in statistical mechanics
\cite{Duplantier_88,Wu_02}.
Like all determinants, those arising in Kirchhoff's theorem
can of course be rewritten as Gaussian integrals
over fermionic (Grassmann) variables.

In this Letter we prove a generalization of Kirchhoff's theorem
in which a large class of combinatorial objects are represented
by suitable non-Gaussian Grassmann integrals.
Although these integrals can no longer be calculated in closed form,
our identities allow the use of field-theoretic methods
to shed new light on the critical behavior of
the underlying geometrical models.

As a special case, we show that unrooted spanning forests,
which arise as a $q \to 0$ limit of the $q$-state Potts model \cite{forests},
can be represented by a Grassmann theory involving a Gaussian term
and a particular bilocal four-fermion term.
Furthermore, this latter model can be mapped,
to all orders in perturbation theory,
onto the $N$-vector model [$O(N)$-invariant $\sigma$-model] at $N=-1$
or, equivalently,
onto the $\sigma$-model taking values in the unit supersphere
in $\R^{1|2}$ [$OSP(1|2)$-invariant $\sigma$-model].
It follows that, in two dimensions,
this fermionic model is perturbatively asymptotically free,
in close analogy to (large classes of) two-dimensional $\sigma$-models
and four-dimensional nonabelian gauge theories.
Indeed, this fermionic model may, because of its great simplicity,
be the most viable candidate for a rigorous
nonperturbative proof of asymptotic freedom ---
a goal that has heretofore remained elusive in both
$\sigma$-models and gauge theories.

The plan of this Letter is as follows:
First we prove some combinatorial identities
involving Grassmann integrals,
culminating in our general formula \reff{eq.genfun2},
and show how a special case yields unrooted spanning forests.
Next we show that this latter model can be mapped onto the
$N$-vector model at $N=-1$, and use this fact to deduce its
renormalization-group (RG) flow at weak coupling.
Finally, we conjecture the nonperturbative phase diagram
in this model.

\paragraph{Combinatorial Identities.}

Let $G=(V,E)$ be a finite undirected graph with vertex set $V$
and edge set $E$.  Associate to each edge $e$ a weight $w_e$,
which can be a real or complex number or, more generally,
a formal algebraic variable.
For $i \neq j$,
let $w_{ij} = w_{ji}$ be the sum of $w_e$ over all edges $e$
that connect $i$ to $j$.
The (weighted) Laplacian matrix $L$ for the graph $G$
is then defined by $L_{ij} = -w_{ij}$ for $i \neq j$,
and $L_{ii} = \sum_{k \neq i} w_{ik}$.
This is a symmetric matrix with all row and column sums equal to zero.

Since $L$ annihilates the vector with all entries 1,
its determinant is zero.
Kirchhoff's matrix-tree theorem \cite{Kirchhoff_et_al}
and its generalizations \cite{Chaiken_et_al}
express determinants of square submatrices of $L$
as generating polynomials of spanning trees
or rooted spanning forests in $G$.
For any vertex $i \in V$,
let $L(i)$ be the matrix obtained from $L$ by deleting
the $i$th row and column.
Then Kirchhoff's theorem states that $\det L(i)$ is independent of $i$
and equals
\be
   \det L(i)  \;=\; \sum_{T \in {\cal T}} \, \prod_{e \in T} w_e   \;,
 \label{eq.Kirchhoff}
\ee
where the sum runs over all spanning trees $T$ in $G$.
(We recall that a subgraph of $G$ is called a tree if it
 is connected and contains no cycles,
 and is called spanning if its vertex set is exactly $V$.)
The $i$-independence of $\det L(i)$ expresses,
in electrical-circuit language,
that it is physically irrelevant which vertex $i$
is chosen to be ``ground''.
There are many different proofs of Kirchhoff's formula \reff{eq.Kirchhoff};
one simple proof is based on the Cauchy--Binet theorem
in matrix theory (see e.g.\ \cite{Moon_70}).

More generally, for any sets of vertices $I,J \subseteq V$,
let $L(I|J)$ be the matrix obtained from $L$
by deleting the columns $I$ and the rows $J$;
when $I=J$, we write simply $L(I)$.
The ``principal-minors matrix-tree theorem'' reads
\be
   \det L(i_1,\ldots,i_r)  \;=\;
   \sum_{F \in {\cal F}(i_1,\ldots,i_r)} \, \prod_{e \in F} w_e   \;,
 \label{eq.principal}
\ee
where the sum runs over all spanning forests $F$ in $G$
composed of $r$ disjoint trees, each of which contains exactly one
of the ``root'' vertices $i_1,\ldots,i_r$.
This theorem can easily be derived by applying
Kirchhoff's theorem \reff{eq.Kirchhoff} to the
graph in which the vertices $i_1,\ldots,i_r$
are contracted to a single vertex.
Finally, the ``all-minors matrix-tree theorem''
(whose proof is more difficult, see \cite{Chaiken_et_al})
states that for any subsets $I,J$ of the same cardinality $r$,
\be
   \det L(I|J)  \;=\;
   \sum_{F \in {\cal F}(I|J)} \epsilon(F,I,J) \, \prod_{e \in F} w_e   \;,
 \label{eq.all-minors}
\ee
where the sum runs over all spanning forests $F$ in $G$
composed of $r$ disjoint trees, each of which contains exactly one
vertex from $I$ and exactly one vertex (possibly the same one) from $J$;
here $\epsilon(F,I,J) = \pm 1$ are signs
whose precise definition is not needed here.

Let us now introduce, at each vertex $i \in V$,
a pair of Grassmann variables $\psi_i$, $\psibar_i$.
All of these variables are nilpotent ($\psi_i^2 = \psibar_i^2 = 0$),
anticommute, and obey the usual rules for Grassmann integration
\cite{Zinn-Justin}.
Writing $\scrd(\psi,\psibar) = \prod_{i \in V} d\psi_i \, d\psibar_i$,
we have, for any matrix $A$,
\be
   \int \scrd(\psi,\psibar) \; e^{\psibar A \psi}
   \;=\;
   \det A
\ee
and more generally
\begin{eqnarray}
   & &  \!\!\!\!\!\!\!\!
        \int \! \scrd(\psi,\psibar) \; \psibar_{i_1} \psi_{j_1} \,\cdots\,
               \psibar_{i_r} \psi_{j_r} \, e^{\psibar A \psi}
     \nonumber \\
   & &  \!\!\!\!\!\!=\,
   \epsilon(i_1,\ldots,i_r|j_1,\ldots,j_r) \,
   \det A(i_1,\ldots,i_r|j_1,\ldots,j_r)
   \;\;
\end{eqnarray}
where the sign $\epsilon(i_1,\ldots,i_r|j_1,\ldots,j_r) = \pm 1$
depends on how the vertices are ordered
but is always $+1$ when $(i_1,\ldots,i_r) = (j_1,\ldots,j_r)$.
These formulae allow us to rewrite the matrix-tree theorems in Grassmann form;
for instance, \reff{eq.principal} becomes
\be
   \int \! \scrd(\psi,\psibar)
               \left( \prod_{\alpha=1}^r \psibar_{i_\alpha} \psi_{i_\alpha}
               \!\right)
               \, e^{\psibar L \psi}
   \;=
   \sum_{F \in {\cal F}(i_1,\ldots,i_r)} \, \prod_{e \in F} w_e   \;.
 \label{eq.principal.2}
\ee

Let us now introduce,
for each connected (not necessarily spanning) subgraph
$\Gamma = (V_\Gamma, E_\Gamma)$ of $G$,
the operator
\be
   Q_\Gamma  \;=\;
   \left( \prod_{e \in E_\Gamma} w_e \right)
   \left( \prod_{i \in V_\Gamma} \psibar_i \psi_i \right)
   \;.
\ee
(Note that each $Q_\Gamma$ is even and hence commutes with
the entire Grassmann algebra.)
Now consider an unordered family $\bGamma = \{\Gamma_1,\ldots,\Gamma_l\}$
with $l \ge 0$,
and let us try to evaluate an expression of the form
\be
   \int \scrd(\psi,\psibar)
               \; Q_{\Gamma_1} \cdots Q_{\Gamma_l}
               \; e^{\psibar L \psi}
   \;.
\ee
If the subgraphs $\Gamma_1,\ldots,\Gamma_l$
have one or more vertices in common,
then this integral vanishes on account of the nilpotency of the
Grassmann variables.
If, by contrast, the $\Gamma_1,\ldots,\Gamma_l$
are vertex-disjoint,
then \reff{eq.principal.2} expresses
$\int \! \scrd(\psi,\psibar) \,
 \left( \prod_{k=1}^l \prod_{i \in V_{\Gamma_k}} \psibar_i \psi_i \right)
 \, e^{\psibar L \psi}$
as a sum over forests rooted at the vertices of
$V_{\bGamma} = \bigcup_{k=1}^l V_{\Gamma_k}$.
In particular, all the edges of
$E_{\bGamma} = \bigcup_{k=1}^l E_{\Gamma_k}$
must be absent from these forests,
since otherwise two or more of the root vertices would lie in the
same component (or one of the root vertices would be connected to itself
by a loop edge).
On the other hand, by adjoining the edges of $E_{\bGamma}$,
these forests can be put into one-to-one correspondence
with what we shall call $\bGamma$-forests,
namely, spanning subgraphs $H$ in $G$ whose edge set contains $E_{\bGamma}$
and which, after deletion of the edges in $E_{\bGamma}$,
leaves a forest in which each tree component contains exactly one
vertex from $V_{\bGamma}$.
(Equivalently, a $\bGamma$-forest is a subgraph $H$
 with $l$ connected components in which each component
 contains exactly one $\Gamma_i$, and which does not contain any cycles
 other than those lying entirely within the $\Gamma_i$.
 Note, in particular, that a $\bGamma$-forest is a forest
 if and only if all the $\Gamma_i$ are trees.)
Furthermore, adjoining the edges of $E_{\bGamma}$
provides precisely the factor $\prod_{e \in E_\bGamma} w_e$.
Therefore
\be
   \int \! \scrd(\psi,\psibar)
               \; Q_{\Gamma_1} \cdots Q_{\Gamma_l}
               \, e^{\psibar L \psi}
   \;=\;
   \sum_{H \in {\cal F}_\bGamma} \, \prod_{e \in H} w_e
 \label{eq.gamma}
\ee
where the sum runs over all $\bGamma$-forests $H$.

We can now combine all the formulae \reff{eq.gamma} into a single
generating function, by introducing a coupling constant $t_\Gamma$
for each connected subgraph $\Gamma$ of $G$.
Since $1 + t_\Gamma Q_\Gamma = e^{t_\Gamma Q_\Gamma}$, we have
\be
   \int \! \scrd(\psi,\psibar)
            \; e^{\psibar L \psi + \sum\limits_\Gamma t_\Gamma Q_\Gamma}
= \!\!\!\!
       \sum_{\begin{scarray}
                \hbox{\scriptsize $\bGamma$ vertex-} \\
                \hbox{\scriptsize disjoint}
             \end{scarray}}
\!\!\!
       \Big( \prod_{\Gamma \in \bGamma} t_\Gamma \Big)
       \sum_{H \in {\cal F}_\bGamma} \, \prod_{e \in H} w_e
   \,.
 \label{eq.genfun1}
\ee
We can express this in another way by interchanging
the summations over $\bGamma$ and $H$.
Consider an arbitrary spanning subgraph $H$
with connected components $H_1,\ldots,H_l$;
let us say that $\Gamma$ {\em marks}\/ $H_i$ (denoted $\Gamma \prec H_i$)
in case $H_i$ contains $\Gamma$ and contains no cycles other than those
lying entirely within $\Gamma$.  Define the weight
\be
   W(H_i)  \;=\; \sum_{\Gamma \prec H_i} t_\Gamma   \;.
\ee
Then saying that $H$ is a $\bGamma$-forest
is equivalent to saying that each of its components
is marked by exactly one of the $\Gamma_i$;
summing over the possible families $\bGamma$, we obtain
\begin{eqnarray}
   & & \!\!\!\!\!\!
        \int \! \scrd(\psi,\psibar)
            \; e^{\psibar L \psi + \sum\limits_\Gamma t_\Gamma Q_\Gamma}
       \nonumber \\
   & & = \!\!
       \sum_{\begin{scarray}
                H \, \hbox{\scriptsize spanning} \subseteq G \\
                H = (H_1,\ldots,H_l)
             \end{scarray}}
       \!\!
       \left( \prod_{i=1}^l W(H_i) \!\right)
       \prod_{e \in H} w_e
       \,.
 \label{eq.genfun2}
\end{eqnarray}
This is our general combinatorial formula.
Extensions allowing prefactors
$\psibar_{i_1} \psi_{j_1} \cdots \psibar_{i_r} \psi_{j_r}$
are also easily derived.

We shall discuss elsewhere some of the applications
of \reff{eq.genfun2}, and restrict attention here to the special case in
which $t_\Gamma = t$ whenever
$\Gamma$ consists of a single vertex with no edges,
$t_\Gamma = u$ whenever $\Gamma$ consists of two vertices
linked by a single edge,
and $t_\Gamma = 0$ otherwise.
We have
\begin{multline}
     \int \! \scrd(\psi,\psibar)
        \, \exp\!\Big[
          \psibar L \psi \,
         +\, t \sum\limits_i \psibar_i \psi_i\,
         +\, u \sum\limits_{\< ij \>} w_{ij} \psibar_i \psi_i \psibar_j \psi_j
               \Big]
\\
   = \!\!\!\!\!\!\!\!
       \sum_{\begin{scarray}
                F \in {\cal F} \\
                F = (F_1,\ldots,F_l)
             \end{scarray}}
       \!\!\!\!\!\!
       \left( \prod_{i=1}^l \,  (t|V_{F_i}| + u|E_{F_i}|) \!\right)
       \,
       \prod_{e \in F} w_e
 \label{eq.fourfermion}
\end{multline}
where the sum runs over spanning forests $F$ in $G$
with components $F_1,\ldots,F_l$;
here $|V_{F_i}|$ and $|E_{F_i}|$ are, respectively,
the numbers of vertices and edges in the tree $F_i$.
[We remark that the four-fermion term
 $u \sum_{\< ij \>} w_{ij} \psibar_i \psi_i \psibar_j \psi_j$
 can equivalently be written, using nilpotency of the Grassmann variables,
 as $-(u/2) \sum_{i,j} \psibar_i \psi_i L_{ij} \psibar_j \psi_j$.]
If $u=0$, this formula represents vertex-weighted spanning forests
as a massive fermionic free field \cite{Duplantier_88,Biggs_93}.
More interestingly,
since $|V_{F_i}| - |E_{F_i}| = 1$ for each tree $F_i$,
we can take $u=-t$
and obtain the generating function of {\em unrooted}\/ spanning forests
with a weight $t$ for each component.
This is furthermore equivalent to giving each edge $e$ a weight $w_e/t$,
and then multiplying by an overall prefactor $t^{|V|}$.
This fermionic representation of unrooted spanning forests is the translation
to generating functions and Grassmann variables of a little-known
but important paper by Liu and Chow \cite{Liu_81}.

The generating function of unrooted spanning forests
is also of interest because it arises as a $q \to 0$ limit
of the $q$-state Potts model, in which the couplings
$v_e = e^{\beta J_e} - 1$ tend to zero with fixed ratios $w_e = v_e/q$
\cite{forests}.

\paragraph{Mapping onto Lattice $\sigma$-Models.}

We now claim that the model \reff{eq.fourfermion} with $u=-t$
can be mapped, to all orders in perturbation theory,
onto the $N$-vector model at $N=-1$.
Recall that the $N$-vector model consists of spins $\bsigma_i \in \R^N$,
$|\bsigma_i| = 1$, located at the sites $i \in V$,
with Boltzmann weight $e^{-\scrh}$ where
$\scrh = -T^{-1} \sum_{\<ij\>} w_{ij} (\bsigma_i \cdot \bsigma_j - 1)$
and $T=\hbox{temperature}$.
Low-temperature perturbation theory is obtained by writing
$\bsigma_i = (\sqrt{1-T\bpi_i^2}, T^{1/2} \bpi_i)$
with $\bpi_i \in \R^{N-1}$
and expanding in powers of $\bpi$.
Taking into account the Jacobian,
the Boltzmann weight is $e^{-\scrh'}$ where
\begin{eqnarray}
   \scrh' & = & \scrh \,+\, \frac{1}{2} \sum\limits_i \log(1-T\bpi_i^2)
       \nonumber \\
   & = &
   \frac{1}{2} \sum\limits_{i,j} L_{ij} \bpi_i \cdot \bpi_j
   \,-\, \frac{T}{2} \sum\limits_i \bpi_i^2
   \,-\, \frac{T}{4} \sum\limits_{\<ij\>} w_{ij} \bpi_i^2 \bpi_j^2
       \nonumber \\
   &  &
   \quad +\,  O(\bpi_i^4, \bpi_j^4)
   \;.
\end{eqnarray}
When $N=-1$, the bosonic field $\bpi$ has $-2$ components,
and so can be replaced by a fermion pair $\psi,\psibar$
if we make the substitution
$\bpi_i \cdot \bpi_j \to \psi_i \psibar_j - \psibar_i \psi_j$.
Higher powers of $\bpi_i^2$ vanish due to the nilpotence of the
Grassmann fields,
and we obtain the model \reff{eq.fourfermion}
if we identify $t=-T$, $u=T$.
Note the reversed sign of the coupling:
the spanning-forest model with positive weights ($t > 0$)
corresponds to the {\em anti}\/ferromagnetic $N$-vector model ($T < 0$).

An alternate mapping can be obtained by introducing at each site,
in addition to the Grassmann fields $\psi_i,\psibar_i$,
an auxiliary one-component bosonic field $\varphi_i$
satisfying the constraint $\varphi_i^2 + 2t \psibar_i \psi_i = 1$.
Solving this constraint yields
$\varphi_i = 1 - t \psibar_i \psi_i = e^{-t\psibar_i \psi_i}$ and
\begin{eqnarray}
   \!\!\!\!\!\!
   \delta(\varphi_i^2 + 2t \psibar_i \psi_i - 1)
   & = &
   \frac{1}{2\varphi_i} \;
   \delta\bigl( \varphi_i - (1 - t \psibar_i \psi_i) \bigr)
      \nonumber \\[1mm]
   & = &
   \frac{e^{t\psibar_i \psi_i}}{2} \,
   \delta\bigl( \varphi_i - (1 - t \psibar_i \psi_i) \bigr)
   .\;\;
\end{eqnarray}
If we define the superfield
$\vecbsigma_i = (\varphi_i, \psi_i, \psibar_i)$
with inner product
$\vecbsigma_i \cdot \vecbsigma_j =
 \varphi_i\varphi_j + t(\psibar_i \psi_j - \psi_i \psibar_j)$,
then the $\sigma$-model
with Hamiltonian
$\scrh = -T^{-1} \sum_{\<ij\>} w_{ij} (\vecbsigma_i \cdot \vecbsigma_j - 1)$
and constraint $\vecbsigma_i \cdot \vecbsigma_i = 1$
corresponds to the fermionic model \reff{eq.fourfermion}
if we again make the identification $t=-T$, $u=T$.
This $\sigma$-model,
which is invariant under the supergroup $OSP(1|2)$,
has been studied previously by one of us \cite{Saleur_and_friends}.
It is presumably nonperturbatively equivalent to the
$N$-vector model at $N=-1$,
on the grounds that each fermion equals $-1$ boson.

It is worth mentioning that the correspondence between the
spanning-forest model and these two $\sigma$-models,
while valid at all orders of perturbation theory,
does {\em not}\/ hold nonperturbatively.
(This can be checked explicitly in the exact solution
for the two-site model \cite{in_prep}.)
The error arises from neglecting the second square root
when solving the constraints;
we did not, in fact, parametrize a (super)sphere
but rather a (super)hemisphere.
Indeed, since $t > 0$ corresponds to an antiferromagnetic $\sigma$-model,
the terms we have neglected are actually dominant!
But no matter: the perturbative correspondence is still correct,
and has the renormalization-group consequences discussed below.
Furthermore, we conjecture that a nonperturbative correspondence
can be obtained by using a $\sigma$-model with
a suitable variant Boltzmann weight.

\paragraph{Continuum Limit.}

Suppose now that the graph $G$ is a regular two-dimensional lattice,
with weight $w_{ij} = w > 0$ for each nearest-neighbor pair.
We can then read off,
from known results on the $N$-vector model \cite{4-loop_all},
the RG flow for the spanning-forest model:
it is
\be
   \frac{d\bar{t}}{d\ell}  \;=\;
   \frac{3}{2\pi} \bar{t}^2 - \frac{3}{(2\pi)^2} \bar{t}^3 +
   \frac{2.34278457}{(2\pi)^3} \bar{t}^4 + \frac{1.43677}{(2\pi)^4} \bar{t}^5 +
   \ldots
 \label{eq.RG}
\ee
where $\bar{t} = t/w$ and
$\ell$ is the logarithm of the length rescaling factor;
here the first two coefficients are universal
(after suitable normalization of the kinetic term),
while the remaining coefficients are for the square lattice only.
The positive coefficient of the $\bar{t}^2$ term
indicates that for $t>0$ the model is perturbatively asymptotically free.
Indeed, two-dimensional $N$-vector models are asymptotically free
for the usual sign of the coupling ($T>0$) when $N > 2$,
but for the reversed sign of the coupling ($T<0$) when $N < 2$.
Assuming that the asymptotic freedom holds also nonperturbatively,
we conclude that for $t>0$ the model is attracted to the
infinite-temperature fixed point at $t=+\infty$,
hence is massive and $OSP(1|2)$-symmetric.
For $t_{\rm crit} < t < 0$, by contrast,
the model is attracted to the free-fermion fixed point at $t=0$,
and hence is massless with central charge $c=-2$,
with the $OSP(1|2)$ symmetry spontaneously broken.
Finally, for $t < t_{\rm crit}$ we expect that the model
will again be massive, with the $OSP(1|2)$ symmetry restored.

More specifically, for $t>0$ it is predicted
that the correlation length diverges for $t \downarrow 0$
(or $w \uparrow +\infty$) as
\begin{eqnarray}
   \!\!\!\!\!\!
   \xi  & = &   C_\xi \,  e^{(2\pi/3)(w/t)} \,
                \left( \frac{2\pi}{3} \frac{w}{t} \right)^{\! 1/3} \;\times
     \nonumber \\
   & &
   \!\! \left[ 1 - 0.0116221204 \frac{t}{w} + 0.00446142 \frac{t^2}{w^2}
                + \ldots \right]
   \;\;
 \label{eq.xi}
\end{eqnarray}
where $C_\xi$ is a nonperturbative constant
(the terms in brackets are for the square lattice only).
The numerical results of \cite{forests},
based on transfer matrices and finite-size scaling,
are consistent with the nonperturbative validity of the
asymptotic-freedom predictions \reff{eq.RG}/\reff{eq.xi},
but are inconclusive because the strip widths are small ($L \le 10$).
It would be interesting to make a Monte Carlo test of \reff{eq.xi},
at large correlation lengths, along the lines of \cite{o3}.

The numerics of \cite{forests}
are also consistent with the central charge $c=-2$ in the
massless phase $t_{\rm crit} < t < 0$,
but are not definitive because of the strong ($1/\log$)
corrections to scaling induced by the marginally irrelevant operator.

Finally, the critical point $t_{\rm crit}$ presumably corresponds
to the $q \to 0$ limit of the antiferromagnetic critical curve
in the $q$-state Potts model, under the identification
$w/t = (e^{\beta J} - 1)/q$.
Known exact results for the square lattice \cite{Baxter_82,Saleur_91}
yield $(w/t)_{\rm crit} = -1/4$.
The analysis of the critical theory proves rather difficult,
but there are strong indications that it is simply a free
$OSP(1|2)$ model, i.e., the theory of a non-compact boson
and a pair of fermions, with central charge $c=-1$.

Let us also remark that there exists a much-studied variant of
the $N$-vector model in which the high-temperature expansion
on the lattice has been truncated so as to forbid loop crossings
\cite{Nienhuis_82}.
For $-2 < N < 2$ this model possesses several critical points;
in particular, the dilute-loop critical point is expected to be generic
in the sense that adding loop crossings acts as an irrelevant perturbation.
For $N=-1$ this yields a $c=-3/5$ theory \cite{central_charge_ON_model};
the relation to the $c=-1$ theory discussed above
is mysterious and deserves further study.

%

It would also be interesting to know whether our identities
are in any way related to the forest-root formula of
Brydges and Imbrie \cite{Brydges-Imbrie},
which leads to a dimensional-reduction formula for branched polymers.

\begin{acknowledgments}
We thank Abdelmalek Abdesselam, David Brydges, John Imbrie,
Marco Polin, Jes\'us Salas and Dominic Welsh for helpful discussions.
This work was supported in part by NSF grant PHY--0099393
and by the DOE.
\end{acknowledgments}


\end{document}